\documentclass{amsart}

\newtheorem{theorem}{Theorem}[section]
\newtheorem{lemma}[theorem]{Lemma}
\newtheorem{corollary}[theorem]{Corollary}
\newtheorem{proposition}[theorem]{Proposition}
\theoremstyle{definition}
\newtheorem{definition}[theorem]{Definition}

\theoremstyle{remark}
\newtheorem{remark}[theorem]{Remark}

\newcommand{\cB}{{\mathcal B}}
\newcommand{\cL}{{\mathcal L}}
\newcommand{\cR}{{\mathcal R}}
\newcommand{\cM}{{\mathcal M}}
\newcommand{\cP}{{\mathcal P}}
\newcommand{\cS}{{\mathcal S}}
\newcommand{\cT}{{\mathcal T}}

\newcommand{\Jm}{J_{\mathrm{m}}}

\newcommand{\bC}{{\mathbb{C}}}
\newcommand{\bN}{{\mathbb{N}}}

\numberwithin{equation}{section}

\newcommand{\Tr}{\mathrm{Tr}}
\newcommand{\id}{\mathrm{id}}
\newcommand{\vr}{\varrho}
\newcommand{\rr}{\vr^{1/2}}

\newcommand{\jed}{{\mathbb{I}}}
\newcommand{\rml}{l}
\newcommand{\rmr}{r}
\newcommand{\la}{\langle}
\newcommand{\ra}{\rangle}

\begin{document}
\title{$k$-decomposability of positive maps}\thanks{Supported by 
RTN HPRN-CT-2002-00279}
\author{W{\l}adys{\l}aw A. Majewski}
\address{Institute of Theoretical Physics and Astrophysics\\ Gda{\'n}sk
University\\ Wita Stwo\-sza~57\\ 80-952 Gda{\'n}sk, Poland\\
E-mail: fizwam@univ.gda.pl}
\author{Marcin Marciniak}
\address{Institute of Mathematics\\ Gda{\'n}sk University\\
Wita Stwosza 57\\ 80-952 Gda{\'n}sk, Po\-land\\
E-mail: matmm@univ.gda.pl}

\maketitle

\section{Introduction}
For any $C^*$-algebra $A$ let $A^+$ denote the set of all positive
elements in $A$. A~{\it state} on a~unital $C^*$-algebra $A$ is a
linear functional $\omega:A\to \mathbb{C}$ such that
$\omega(a)\geq 0$ for every $a\in A^+$ and $\omega(\mathbb{I})=1$
where $\mathbb{I}$ is the unit of $A$. By $\cS(A)$ we will denote
the set of all states on $A$. For any Hilbert space $H$ we denote
by $\cB(H)$ the set of all bounded linear operators on $H$.

A linear map $\varphi:A\rightarrow B$ between $C^*$-algebras is called {\it positive}
if $\varphi(A^+)\subset B^+$. For $k\in\mathbb{N}$ we consider a~map
$\varphi_k:M_k(A)\to M_k(B)$ where $M_k(A)$ and $M_k(B)$ are the algebras
of $k\times k$
matrices with coefficients from $A$ and $B$ respectively, and
$\varphi_k([a_{ij}])=[\varphi(a_{ij})]$.
We say that $\varphi$ is {\it $k$-positive}
if the map $\varphi_k$ is positive. The map $\varphi$ is said to be
{\it completely positive}
when it is $k$-positive for every $k\in\mathbb{N}$.

A {\it Jordan morphism} between $C^*$-algebras $A$ and $B$ is
a~linear map $\rho:A\to B$ which respects the Jordan structures
of algebras $A$ and $B$, i.e. $\rho(ab+ba)= \rho(a)\rho(b)+
\rho(b)\rho(a)$ for every $a,b\in A$. Let us recall that every
Jordan morphism is a positive map but it need not be a completely
positive one (in fact it need not even be 2-po\-si\-ti\-ve). It is
commonly known (\cite{Sto3}) that every Jordan morphism $\rho:A\to
\cB(H)$ is a sum of a $^*$-morphism and a $^*$-antimorphism.

The Stinespring theorem states that every completely positive map
$\varphi:A\to \cB(H)$ has the form $\varphi(a)=W^*\pi(a)W$, where
$\pi$ is a $^*$-representation of $A$ on some Hilbert space $K$,
and $W$ is a bounded operator from $H$ to $K$.

Following St{\o}rmer (\cite{Sto1}) we say that a map
$\varphi:A\to \cB(H)$ is {\it decomposable} if there are a Hilbert space $K$,
a Jordan morphism $\rho:A\to \cB(K)$, and a bounded linear operator
$W$ from $H$ to $K$ such that $\varphi(a)=W^*\rho(a)W$ for every $a\in A$.

By $B(H)\ni b\mapsto b^t\in B(H)$ we denote the transposition map
(for details see Section 2).
We say that a linear map $\varphi:A\to \cB(H)$ is
\textit{$k$-copositive} (resp. \textit{completely copositive}) if
the map $a\mapsto \varphi(a)^{\mathrm{t}}$ is $k$-positive (resp.
\textit{completely positive}). The following theorem (\cite{Sto2}) characterizes
decomposable maps in the spirit of Stinespring's theorem:
\begin{theorem}
Let $\varphi:A\to \cB(H)$ be a linear map. Then the fo\-llo\-wing conditions
are equivalent:
\begin{enumerate}
\item
$\varphi$ is decomposable;
\item
for every natural number $k$ and for every matrix
$[a_{ij}]
\in M_k(A)$
such that both $[a_{ij}]$ and $[a_{ji}]$ belong to $M_k(A)^+$ the matrix $[\varphi(a_{ij})]$
is in $M_k(\cB(H))^+$;
\item
there are maps $\varphi_1,\varphi_2:A\to \cB(H)$ such that
$\varphi_1$ is completely positive and $\varphi_2$ completely
copositive, with $\varphi=\varphi_1+\varphi_2$.
\end{enumerate}
\end{theorem}

The classification of decomposable maps is still not
complete even in the case when $A$ and $H$ are finite dimensional, i.e.
$A=\cB(\bC^m)$ and $H=\mathbb{C}^n$.
The most important step was done by St{\o}rmer (\cite{Sto2}), Choi (\cite{Ch2,Ch3}) and
Woronowicz (\cite{Wor}).
St{\o}rmer and Woronowicz proved that if $m=n=2$ or $m=2$, $n=3$
then every positive map is decomposable.
The first examples of nondecomposable maps was given by Choi (in the case $m=n=3$)
and Woronowicz (in the case $m=2$, $n=4$).
It seems that very general positive maps (so not of the CP
class) and hence possibly non-decomposable ones, are crucial for
an analysis of nontrivial quantum correlations, i.e. for an
analysis of genuine quantum maps (\cite{Wit,Per,Hor,MajO,MajL,MajC}).
Having that motivation in mind in our last paper (\cite{LMMp}) we presented a step toward a
canonical prescription for the construction of decomposable and
non-decomposable maps. Namely, we studied the notion of
$k$-decomposability and proved an analog of Theorem 1.1. Moreover
it turned out that it is possible to describe the notion of $k$-decomposabilty
in the dual picture. More precisely, the analog of Tomita-Takesaki construction
for the transposition map on the algebra $B(H)$ can be formulated (Section 2). 
Application of this scheme provides us with a new characterization of decomposability
on the Hilbert space level (see Section 3). Thus, it can be said that we are using the equivalence
of the Schr\"odinger and Heisenberg pictures in the sense of
Kadison (\cite{Kad}), Connes (\cite{Conn}) and Alfsen, Shultz (\cite{Alf}).
Section 4 provides a detailed exposition of two dimensional case and establishes the relation between 
St{\o}rmer construction of local decomposability and decomposability fo distinguished subsets of positive maps.

\section{Tomita-Takesaki scheme for transposition}\label{Tomita}
Let $H$ be a finite dimensional (say $n$-dimensional) Hilbert space. 
Define $\omega\in \cB(H)_{+,1}^*$ as
$\omega(a)= \Tr\varrho a,$
where $\varrho$ is an invertible density matrix, i.e. the state $\omega$
is a faithful one.
Denote by $(H_\pi,\pi,\Omega)$ the GNS triple associated with
$(\cB(H),\omega)$.
Then, one can identify the Hilbert space
$H_\pi$ with $\cB(H)$ where the inner product
$(\cdot\,,\cdot)$ defined
as $(a,b)=\Tr a^*b$ for $a,b\in \cB(H)$.
With the above identification one has $\Omega= \rr$ and
$\pi(a)\Omega=a\Omega$ for $a\in B(H)$.
In this setting one can simply express the modular conjugation $\Jm$ 
as the hermitian involution, i.e.
$\Jm a \rr =\rr a^*$ for $a\in B(H)$. 
Similarly,
the modular operator $\Delta$ is equal to the map
$\varrho \cdot\varrho^{-1}$;

Let $\{x_i\}_{i=1,\ldots,n}$ be the orthonormal basis of $H$ consisted of eigenvectors of $\varrho$.
Then we can define 
\begin{equation}\label{Jcdef}
J_c f = \sum_i \overline{\langle x_i,f\rangle} x_i
\end{equation}
for every $f\in H$. The map $J_c$ is a conjugation on $H$. So, we can define
the transposition on $\cB(H)$ as the map $a\mapsto a^t\equiv J_ca^*J_c$ where
$a\in \cB(H)$. By $\tau$ we will denote the map induced on $H_\pi$ by the transposition, i.e.
\begin{equation}\label{tau}
\tau a\rr=a^t\rr
\end{equation}
where $a\in \cB(H)$.

Let $E_{ij}= |x_i\rangle \langle x_j|$ for $i,j=1,\ldots,n$. Obviously, $\{E_{ij}\}$ is an
orthonormal basis in $H_\pi$. Hence, similarly to (\ref{Jcdef}) one can can define
a conjugation $J$ on
$H_\pi$
\begin{equation}\label{Jdef}
J a \rr = \sum_{ij} \overline{(E_{ij},a \rr)} E_{ij}
\end{equation}
where $a\in B(H)$. We have the following
\begin{proposition}\label{transposition}
Let $a\in \cB(H)$ and $\xi\in H_\pi$. Then
$$a^t\xi=Ja^*J\xi.$$
\end{proposition}

Now, define the unitary operator $U$ on $H_\pi$ by
\begin{equation}\label{U}
U = \sum_{ij} |E_{ji}\rangle \langle E_{ij}|
\end{equation}
Clearly, $UE_{ij} = E_{ji}$. The properties of $U$, introduced above conjugation $J$ and
modular conjugation $J_m$ is are described by
\begin{proposition}\label{commute}
One has:
\begin{enumerate}
\item\label{commute1} $U^2 = \jed$ and $U = U^*$
\item\label{commute2} $J=U\Jm$;
\item\label{commute3} $J$, $\Jm$ and $U$ mutually commute;
\item $J$ commutes with the modular operator $\Delta$.
\end{enumerate}
\end{proposition}

The following theorem justifies using the term "Tomita-Takesaki scheme" for transposition
\begin{theorem}\label{polar}
If $\tau$ is the map introduced in (\ref{tau}), then
$$\tau=U\Delta^{1/2}.$$
Moreover one has the following properties:
\begin{enumerate}
\item $U\Delta=\Delta^{-1}U$;
\item\label{polar2} If $\alpha$ is the automorphism of $B(H_\pi)$ implemented by $U$, i.e.
$\alpha(x)=UxU^*$ for $x\in B(H_\pi)$, then $\alpha$ maps $\pi(B(H))$ onto its commutant
$\pi(B(H))^\prime$;
\item If $V_\beta$ denotes the cone 
$\overline{\left\{\Delta^\beta a\varrho^{1/2}:\,a\in B(H)^+\right\}}\subset H_\pi$ (cf.  \cite{Ara}) 
for each $\beta\in [0,1/2]$
then $U$ maps $V_\beta$ onto $V_{(1/2)-\beta}$. In particular, the natural cone $\cP=V_{1/4}$
is invariant with respect to $U$.
\end{enumerate}
\end{theorem}

\begin{corollary}
$U \Delta^{1/2}$ 
maps $V_0$ into itself.
\end{corollary}

Summarizing, this section establishes a close relationship between
the Tomita-Takesaki scheme and transposition.
Moreover, we have the following :
\begin{proposition}\label{transofstates}
Let $\xi\mapsto\omega_\xi$ be the homeomorphism (\cite{Conn,BR}) between the natural
cone $\cP$ and the set of normal states on $\pi(\cB(H))$ such that
$$\omega_\xi(a)=(\xi,a\xi),\quad a\in \cB(H).$$
For every state $\omega$ define
$\omega^\tau(a)=\omega(a^t)$ where $a\in \cB(H)$.
If $\xi\in\cP$ then the unique vector in $\cP$ mapped 
into the state
$\omega_\xi^\tau$ by the homeomorphism described above, 
is equal to $U\xi$
\end{proposition}

In the sequel, we will need the following construction:
Suppose that we have a $C^*$-algebra $A$ equipped with
a faithful state $\omega_A$ and consider the tensor
product $A\otimes B(H)$, where $H$ is the same as above.
Then $\omega_A\otimes\omega$ is a faithful state on $A\otimes B(H)$.
So, we can perform GNS constructions for both $(A,\omega_A)$ and
$(A\otimes B(H),\omega_A\otimes\omega)$ and obtain
representations $(H_A,\pi_A,\Omega_A)$ and $(H_\otimes,\pi_\otimes,\Omega_\otimes)$
respectively.
We observe that we can make the following identifications:
\begin{enumerate}
\item
$H_\otimes=H_A\otimes H_\pi$,
\item
$\pi_\otimes=\pi_A\otimes\pi$,
\item
$\Omega_\otimes=\Omega_A\otimes\Omega$
\end{enumerate}
where $(H_\pi,\pi,\Omega)$ is the GNS triple described in the begining of this section.
With these identifications we have $J_\otimes=J_A\otimes J_m$ and $\Delta_\otimes=\Delta_A\otimes\Delta$
where $J_\otimes$, $J_A$, $J_m$ are modular conjugations and $\Delta_\otimes$, $\Delta_A$, $\Delta$
are modular operators for $(\pi_\otimes(A\otimes B(H))'',\Omega_\otimes)$,
$(\pi_A(A)'',\Omega_A)$, $(\pi(B(H))'',\Omega)$ respectively. Since $\Omega_A$ and $\Omega$
are separating vectors, we will write $a\Omega_A$ and $b\Omega$ instead of $\pi_A(a)\Omega_A$
and $\pi(b)\Omega$ for $a\in A$ and $b\in B(H)$.

The natural cone (\cite{BR,Ara}) $\cP_\otimes$ for $(\pi_\otimes(A\otimes B(H))'',\Omega_\otimes)$ is defined
as the closure of the set
$$\left\{\left(\sum_{k=1}^m a_k\otimes b_k\right)j_\otimes\left(\sum_{l=1}^m a_l\otimes
b_l\right)\Omega_\otimes:\,\begin{array}{l}
{m\in\bN} \\ a_1,\ldots,a_m\in A \\ b_1,\ldots,b_m\in B(H)\end{array}\right\}$$
where $j_\otimes(\cdot)=J_\otimes\cdot J_\otimes$ is the modular morphism 
on $\pi_\otimes(A\otimes B(H))''=\pi_A(A)''\otimes \pi(B(H))''$.

Motivated by Proposition \ref{transofstates} we introduce the 
"transposed cone" $\cP_\otimes^\tau=(\jed\otimes U)\cP_\otimes$ where
$U$ is the unitary defined in (\ref{U}). Elements of this cone are in 1-1 correspondence
with the set of partial transpositions $\phi\circ(\id\otimes t)$ for all states $\phi$ on $A\otimes B(H)$.
It can be easily calculated
that we have the following 
\begin{theorem}
The transposed cone $\cP_\otimes^\tau$ is the closure of the set
$$\left\{\left(\sum_{k=1}^m a_k\otimes \alpha(b_k)\right)j_\otimes\left(\sum_{l=1}^m a_l\otimes
\alpha(b_l)\right)\Omega_\otimes:\,
\begin{array}{l}m\in\bN \\ a_1,\ldots,a_m\in A \\ b_1,\ldots,b_m\in B(H)\end{array}\right\}$$
where $\alpha$ is the automorphism introduced in Theorem \ref{polar}(\ref{polar2}).

Consequently, $\cP_\otimes^\tau=\cP_\otimes'$ where $\cP_\otimes'$ is the
natural cone for $(\pi_A(A)\otimes\pi(B(H))',\omega_\otimes)$.
\end{theorem}

\section{$k$-decomposability at the Hilbert-space level}\label{hilbert}
Let $A$ be a unital $C^*$-algebra, $H$ be a Hilbert space and let
$\varphi:A\longrightarrow B(H)$ be a linear map. We introduce (\cite{LMMp})
the notion of $k$-decomposability of the map $\varphi$ was studied.
\begin{definition}
\begin{enumerate}
\item
We say that $\varphi$ is {\em $k$-decomposable}
if there are maps $\varphi_1,\varphi_2:A\to \cB(H)$ such that $\varphi_1$
is $k$-positive, $\varphi_2$ is $k$-copositive and $\varphi=\varphi_1+\varphi_2$.
\item
We say that $\varphi$ is {\em weakly $k$-decomposable} if there is a $C^*$-algebra $E$,
a unital Jordan morphism $\rho:A\to E$, and a positive map
$\psi :E\to \cB(H)$ such that $\psi|_{\rho(A)}$ is $k$-positive and
$\varphi=\psi\circ \rho$.
\end{enumerate}
\end{definition}
The connection between $k$-decomposability, weak $k$-decomposability and the St{\o}rmer
condition (\cite{Sto1}) is the following
\begin{theorem}\label{k-dec}
For any linear map $\varphi:A\to \cB(H)$ consider the following conditions:
\begin{itemize}
\item[{\rm (D$_k$)}]
$\varphi$ is $k$-decomposable;
\item[{\rm (W$_k$)}]
$\varphi$ is weakly $k$-decomposable;
\item[{\rm (S$_k$)}]
for every matrix $[a_{ij}]\in M_k(A)$ such that both $[a_{ij}]$ and $[a_{ji}]$ are in
$M_k(A)^+$ the matrix $[\varphi(a_{ij})]$ is positive in $M_k(\cB(H))$;
\end{itemize}
Then we have the following implications:
{\rm (D$_k$)} $\Rightarrow$ {\rm (W$_k$)} 
$\Leftrightarrow$ {\rm (S$_k$)}.
\end{theorem}

The results of Section \ref{Tomita} strongly suggest that a more
complete theory of $k$-decomposable maps may be obtained in
Hilbert-space terms. To examine that question we will study the
description of positivity in the dual approach to that given in
in the above theorem, i.e. we will be concerned with the approach on
the Hilbert space level.

Let us resrict to the case $\varphi:{\mathcal M}\longrightarrow {\mathcal M}$ where
$\mathcal M \subset \cB(H_\cM)$ is a concrete von Neumann algebra
with a cyclic and separating vector $\Omega_\cM$. When used, $\omega_\cM$
will denote the vector state $\omega_\cM(\cdot) = (\Omega_\cM, \cdot\Omega_\cM)$. The
natural cone (modular operator) associated with $({\mathcal M},
\Omega_\cM)$ will be denoted by $\cP_\cM$ ($\Delta_\cM$ respectively).

We assume that that $\varphi$ satisfy Detailed Balance II (\cite{MS}), i.e. there is
a positive unital map $\varphi^\beta$ such that $\omega(a^*\varphi(b))=
\omega(\varphi^\beta(a^*)b)$ for $a,b\in\cM$. In this
case $\varphi$ induces a bounded operator $T_\varphi$ on $H_{\omega_\cM}$ which commutes
strongly with $\Delta_\cM$ and satisfies $T_\varphi^*(\cP_\cM)\subset\cP_\cM$.

Let $B(\bC^n)\ni a\mapsto a^t\in B(\bC^n)$ denotes the usual transposition map
on the algebra of $n\times n$-matrices. Let $\omega$ be a faithful state on
$B(\bC^n)$ and let $\cP_n$ denote the natural cone for $(\cM\otimes B(\bC^n),\omega_\cM\otimes\omega)$.
From Theorem \ref{k-dec} it follows that to develop the theory of decomposability on the Hilbert space level
we should examine the action of the map $\jed\otimes T_\varphi$ on the
transposed cone $\cP_n^\tau=(\jed\otimes U)\cP_n$ described in the previous section, where the operator $U$ on $B(\bC^n)$
was introduced in the previous section (for some orthonormal basis $\{e_i\}$ of
eigenvectors of $\varrho_{\omega_0}$).
It can be deduced from Proposition 2.5.26 in the book of Bratteli and Robinson (\cite{BR}) and the results of previous section that
the cones $\cP_n$ and $\cP_n^\tau$ have the following forms:
$$\mathcal{P}_n = \overline{{\Delta}^{1/4}_n\{[a_{ij}]\Omega_n :
[a_{ij}] \in M_n(\mathcal{M})^+\}},$$ 
$$\mathcal{P}_n^\tau = \overline{{\Delta}^{1/4}_n\{[a_{ji}]\Omega_n :
[a_{ij}] \in M_n(\mathcal{M})^+\}}.$$ 
It turns out that the adaptation of Lemma 4.10 in the paper of Majewski (\cite{Maj})
leads to the following characterization of $k$-positivity and $k$-copositivity
\begin{lemma}
The map $\varphi:\cM\longrightarrow\cM$ is $k$-positive ($k$-copositive) if and only if
$(T_\varphi\otimes\jed)^*(\cP_n)\subset\cP_n$ (respectively $(T_\varphi\otimes\jed)^*(\cP_n)\subset
\cP_n^\tau$) for every $n=1,\ldots,k$.
\end{lemma}
By $\overline{\mathrm co}(\cT)$ we denote the closed convex hall of the subset $\cT$.
Now, we are in position to give promised result.
\begin{theorem}
Consider the following two conditions on $\varphi$:
\begin{enumerate}
\item
$\varphi$ is weakly $k$-decomposable;
\item
$(T_\varphi\otimes\jed)^*(\cP_n)\subset\overline{\mathrm co}(\cP_n\cup\cP_n^\tau)$ for every
$n=1,\ldots,k$.
\end{enumerate}
Then, in general, the property (2) implies (1). If, in addition, the cone $\cP_n\cap\cP_n^\tau$
is equal to the closure of the set
$$\{\Delta_n^{1/4}[a_{ij}]\Omega_n:\,[a_{ij}],[a_{ji}]\in M_n(\cM)^+\}$$
then (2) follows from (1).

In particular in the finite-dimensional case the two conditions are equivalent.
\end{theorem}
\begin{remark}
It can be easily showed that $\overline{\mathrm co}(\cP_n\cup\cP_n^\tau)$ and
$\cP_n\cap\cP_n^\tau$ are dual cones. It is still an open question
whether the equality
$$\overline{\{\Delta_n^{1/4}[a_{ij}]\Omega_n:\,[a_{ij}],[a_{ji}]\in M_n(\cM)^+\}}=
\cP_n\cap\cP_n^\tau$$
holds in general.
\end{remark}

\section{Application of local decomposability to low dimensional cases}
In this section we indicate how the discussed techniques relied on decomposition
may be used for a characterization of linear positive unital maps 
$\varphi:M_m(\bC)\longrightarrow M_n(\bC)$ in the case of low dimensions $m$ and $n$. 

St{\o}rmer (\cite{Sto1}) proved that
each positive map $\varphi:M_m(\bC)\longrightarrow M_n(\bC)$ is locally decomposable, i.e.
for every non-zero vector $\eta\in\bC^n$ there exist a Hilbert space $K_\eta$, a linear map
$V_\eta$ on $K_\eta$ into $\bC^n$, such that $\Vert V_\eta\Vert\leq M$ for all $\eta$, and a
Jordan $^*$-homomorphism $\rho_\eta$ of $M_m(\bC)$ such that
\begin{equation}\label{locdec}
\varphi(a)\eta=V_\eta\rho_\eta(a)V_\eta^*\eta
\end{equation}
for all $a\in M_m(\bC)$.
For the readers convenience we remind the construction of St{\o}rmer in details. Given a vector $\eta\in \bC^m$,
$\Vert\eta\Vert=1$, we consider the state $\omega_\eta$ on $M_n(\bC)$ defined as
\begin{equation}\label{omega}
\omega_\eta(a)=\langle\eta,\varphi(a)\eta\rangle,\;\;\;a\in M_n(\bC).
\end{equation}
Let $\mathcal{L}=\{a\in M_n(\bC):\,\omega_\eta(a^*a)=0\}$ and $\mathcal{R}=\{a\in M_n(\bC):\,\omega_\eta(aa^*)=0\}$.
Observe that $\mathcal{L}$ is a left ideal in $M_n(\bC)$ while $\mathcal{R}$ is a right ideal. By $K_\rml$ and
$K_\rmr$ we denote the quotient spaces $M_n(\bC)/\mathcal{L}$ and $M_n(\bC)/\mathcal{R}$ respectively.
For any $a\in M_n(\bC)$ we write $[a]_\rml$ and $[a]_\rmr$ the abstract classes of $a$ in $K_\rml$ and $K_\rmr$
respectively.
Next, let $K_\eta=K_\rml\oplus K_\rmr$ and define the scalar product $\langle\langle\cdot,\cdot\rangle\rangle$ on $K_\eta$
\begin{equation}\label{product}
\langle\langle[a_1]_\rml \oplus[a_2]_\rmr,[b_1]_\rml \oplus[b_2]_\rmr\rangle\rangle=\frac{1}{2}
\omega_\eta(a_1^*b_1)+\frac{1}{2}\omega_\eta(b_2a_2^*).
\end{equation}
For simplicity we will write $[a]$ instead of $[a]_\rml\oplus[a]_\rmr$ for $a\in M_n(\bC)$. By $G$ we denote the subspace
of $K_\eta$ consisted of every such elements and by $G'$ its orthogonal complement. Finally, $V_\eta$ and $\rho_\eta$ are
given by
\begin{equation}\label{rho}
\rho(a)\left([b_1]_\rml\oplus[b_2]_\rmr\right)=[ab_1]_\rml\oplus[b_2a]_\rmr,\;\;\;a,b_1,b_2\in M_n(\bC);
\end{equation}
\begin{equation}\label{V}
V_\eta k=\left\{\begin{array}{ll}
\varphi(a)\eta,&\mbox{if $k=[a]$ for some $a\in M_n(\bC)$,}\\
0,&\mbox{if $k\in G'$.}
\end{array}\right.
\end{equation}

The crucial point in our considerations is the characterization of face structure of the set of unital positive maps
between matrix algebras which was done by Kye (\cite{Kye}). Recall that if $C$ is a convex set then a convex subset 
$F\subset C$ is called a face if for any $x,y\in C$ and $0<\lambda<1$ the following implication holds: 
$$\lambda x+(1-\lambda)y\in F\;\;\Rightarrow\;\;x,y\in F.$$
Kye proved that any maximal face of the set of unital positive maps from $M_m(\bC)$ into $M_n(\bC)$ is of the following 
form
\begin{equation}\label{face}
F_{\xi,\eta}=\{\varphi:\, \varphi(\jed)=\jed,\,\varphi(|\xi\rangle\langle\xi|)\eta=0\}
\end{equation}
for some $\xi\in\bC^m$ and $\eta\in\bC^n$. 

In the sequel we describe the case $m=n=2$. As we mentioned in the introduction, St{\o}rmer (\cite{Sto1}) proved that
every positive map $\varphi:M_2(\bC)\rightarrow M_2(\bC)$ is (globaly!) decomposable. The next proposition indicates 
the relationship between this phenomenon and the notion of local decomposability for $\varphi$ in a maximal face.
We need the following notations. If $\xi$ and $\eta$ are arbitrary unit vectors in $\bC^2$ then
let $\xi_1,\xi_2$ be an orthonormal basis in $\bC^2$ such that $\xi_1=\xi$ and similarly $\eta_1,\eta_2$ be a basis
such that $\eta_1=\eta$. By $e_{ij}$ we denote the operator $|\xi_i\ra\la\xi_j|$ for $i,j=1,2$.
\begin{proposition}
Suppose $\varphi\in F_{\xi,\eta}$. Let $K_\eta$,
$V_\eta$ and $\rho_\eta$ be as in (\ref{locdec}). Then 
\begin{equation}\label{main}
\varphi(a)=V_\eta\rho_\eta(a)V_\eta^*,\;\;\; a\in M_2(\bC).
\end{equation}
if and only if
\begin{equation}\label{tr}
\Tr\varphi(e_{12})=\Tr\varphi(e_{21})=0,\;\;\;
\Tr\varphi(e_{22})=1,
\end{equation}
\begin{equation}\label{alfabeta}
\Tr\varphi(e_{11})=2\left(|\la\eta_2,\varphi(e_{12})\eta_1\ra|^2+|\la\eta_2,\varphi(e_{21})\eta_1\ra|^2\right).
\end{equation}
\end{proposition}
\begin{proof}
From the definition (\ref{face}) of $F_{\xi,\eta}$ it follows that the projection $e_{11}$ is an element
of $\cL$ and $\cR$. On the other hand both $\cL$ and $\cR$ are proper ideals in $M_2(\bC)$ because $\varphi$
is unital. Consequently $\cL=M_2(\bC)e_{11}$ and $\cR=e_{11}M_2(\bC)$ and the Hilbert space 
$K=M_2(\bC)/M_2(\bC)e_{11}\oplus M_2(\bC)/e_{11}M_2(\bC)$ is four dimensional. By direct computations
it can be checked that the elements
\begin{eqnarray*}
k_1&=&\sqrt{2}[e_{12}]=\sqrt{2}[e_{12}]_\rml+\sqrt{2}[e_{12}]_\rmr\\
k_2&=&\sqrt{2}[e_{21}]=\sqrt{2}[e_{21}]_\rml+\sqrt{2}[e_{21}]_\rmr\\
k_3&=&[e_{22}]=[e_{22}]_\rml+[e_{22}]_\rmr\\
k_4&=&[e_{22}]_\rml-[e_{22}]_\rmr
\end{eqnarray*}
form an orthonormal basis in $K$. Moreover, 
from (\ref{V}) and (\ref{face}) we get the following equalities
$$V_\eta k_1=\sqrt{2}\varphi(e_{12})\eta_1=
\sqrt{2}\la\eta_1,\varphi(e_{12})\eta_1\ra\eta_1+\sqrt{2}\la\eta_2,\varphi(e_{12})\eta_1\ra\eta_2=\alpha\eta_2$$
where $\alpha=\sqrt{2}\la\eta_2,\varphi(e_{12})\eta_1\ra$. The last equality is due to the fact that 
$e_{12}\in e_{11}M_2(\bC)
\subset\ker\omega_\eta$ (cf. (\ref{omega})). Similarly we check that 
$$V_\eta k_2=\beta\eta_2$$
where $\beta=\sqrt{2}\la\eta_2,\varphi(e_{21})\eta_1\ra$. The definition (\ref{face}) of $F_{\xi,\eta}$ and the fact
that $\varphi$ is unital imply
$$V_\eta k_3=\varphi(e_{22})\eta_1=\varphi(\jed)\eta_1-\varphi(e_{11})\eta_1=\eta_1.$$
Finally, from the dafinition (\ref{V}) of $V_\eta$ and from the fact that $k_4$ is orthogonal to $G$ it follows
that
$$V_\eta k_4=0.$$
Hence, the matrix of the operator $V_\eta:K\rightarrow\bC^2$ in the bases $\{k_1,k_2,k_3,k_4\}$ and $\{\eta_1,\eta_2\}$
has the form
\begin{equation}\label{matrix}V_\eta=\left[\begin{array}{cccc}0&0&1&0\\ \alpha&\beta&0&0\end{array}\right].\end{equation}

In order to prove sufficiency in the statement of the theorem, assume that 
$\varphi$ fulfils conditions (\ref{tr}) and (\ref{alfabeta}). We will show that 
$\varphi(a)=V_\eta\rho_\eta(a)V_\eta^*$ for any
$a\in M_2(\bC)$. Observe that from local decomposability (\ref{locdec}) we have
$$\varphi(a)\eta_1=V_\eta\rho_\eta(a)V_\eta^*\eta_1.$$
So, it remains to prove that
\begin{equation}\label{eta2}
\varphi(a)\eta_2=V_\eta\rho_\eta(a)V_\eta^*\eta_2
\end{equation}
for every $a\in M_2(\bC)$. As $\{e_{ij}\}_{i,j=1,2}$ forms a system of matrix units in $M_2(\bC)$ it is enough
to show (\ref{eta2}) for $a=e_{ij}$ where $i,j=1,2$.

Let $a=e_{12}$. Following the assumption and the fact that $e_{12}\in e_{11}M_2(\bC)\subset \ker\omega_\eta$ we have
$$0=\Tr\varphi(e_{12})=\la\eta_1,\varphi(e_{12})\eta_1\ra+\la\eta_2,\varphi(e_{12})\eta_2\ra=
\la\eta_2,\varphi(e_{12})\eta_2\ra.$$
Hence, 
$$\varphi(e_{12})\eta_2=\la\eta_1,\varphi(e_{12})\eta_2\ra\eta_1+\la\eta_2,\varphi(e_{12})\eta_2\ra\eta_2=
\la\eta_1,\varphi(e_{21}\eta_2\ra\eta_1.$$
On the other hand, application of (\ref{matrix}) and (\ref{rho}) yields
\begin{eqnarray*}
\lefteqn{V_\eta\rho(e_{12})V_\eta^*\eta_2=}\\
&=& V_\eta\rho(e_{12})\left(\overline{\alpha}k_1+\overline{\beta}k_2\right)\\
&=& 2V_\eta\left(\la\eta_1,\varphi(e_{21})\eta_2\ra\rho_\eta(e_{12})[e_{12}]+\la\eta_1,\varphi(e_{12})\eta_2\ra
\rho_\eta(e_{12})[e_{21}]\right)\\
&=& 2\la\eta_1,\varphi(e_{12})\eta_2\ra V_\eta\left([e_{11}]_\rml+[e_{22}]_\rmr\right)\\
&=& \la\eta_1,\varphi(e_{12})\eta_2\ra V_\eta\left(k_3-k_4\right)\\
&=& \la\eta_1,\varphi(e_{12})\eta_2\ra \eta_1
\end{eqnarray*}
So, we get $\varphi(e_{12})\eta_2=V_\eta\rho_\eta(e_{12})V_\eta^*\eta_2$.

By simillar computations we check that $\varphi(e_{21})\eta_2=V_\eta\rho(e_{21})V_\eta^*\eta_2$.

Now, let $a=e_{22}$. We have
$$1=\Tr\varphi(e_{22})=\la\eta_1,\varphi(e_{22})\eta_1\ra+\la\eta_2,\varphi(e_{22})\eta_2\ra=
1+\la\eta_2,\varphi(e_{22})\eta_2\ra,$$
so $\la\eta_2,\varphi(e_{22})\eta_2\ra=0$ and consequently
\begin{eqnarray*}
\varphi(e_{22})\eta_2&=&\la\eta_1,\varphi(e_{22})\eta_2\ra\eta_1+\la\eta_2,\varphi(e_{22})\eta_2\ra\eta_2\\
&=&
\la\varphi(e_{22})\eta_1,\eta_2\ra=\la\eta_1,\eta_2\ra=0.
\end{eqnarray*}
Moreover,
$$V_\eta\rho_\eta(e_{22})V_\eta^*\eta_2=
2V_\eta\left(\la\eta_1,\varphi(e_{21})\eta_2\ra [e_{12}]_\rmr+\la\eta_1,\varphi(e_{12})\eta_2\ra [e_{21}]_\rml\right)=0.$$
The last equality follows from the fact that $e_{12}\in e_{11}M_2(\bC)=\cR$ and $e_{21}\in M_2(\bC)e_{11}=\cL$.

Finally, let $a=e_{11}$. Then
\begin{eqnarray*}
\lefteqn{V_\eta\rho_\eta(e_{11})V_\eta^*\eta_2=}\\
&=&V_\eta\left(2\la\eta_1,\varphi(e_{21})\eta_2\ra\rho_\eta(e_{11})[e_{12}]+
2\la\eta_1,\varphi(e_{12})\eta_2\ra\rho_\eta(e_{11})[e_{21}]\right)\\
&=& 2\la\eta_1,\varphi(e_{21})\eta_2\ra V_\eta [e_{12}]_\rml+
2\la\eta_1,\varphi(e_{12})\eta_2\ra V_\eta[e_{21}]_\rmr\\
&=& \sqrt{2}\la\eta_1,\varphi(e_{21})\eta_2\ra V_\eta k_1+
\sqrt{2}\la\eta_1,\varphi(e_{12})\eta_2\ra V_\eta k_2\\
&=& 2\left(|\la\eta_2,\varphi(e_{21})\eta_1\ra|^2+|\la\eta_2,\varphi(e_{12})\eta_1\ra|^2\right)\eta_2.
\end{eqnarray*}
As $\varphi\in F_{\xi,\eta}$ then
$$\Tr\varphi(e_{11})=\la\eta_1,\varphi(e_{11})\eta_1\ra+\la\eta_2,\varphi(e_{11})\eta_2\ra=\la\eta_2,\varphi(e_{11})
\eta_2\ra,$$
hence
$$\varphi(e_{11})\eta_2=\la\eta_1,\varphi(e_{11})\eta_2\ra\eta_1+\la\eta_2,\varphi(e_{11})\eta_2\ra\eta_2=
\la\eta_2,\varphi(e_{11})\eta_2\ra\eta_2=\left[\Tr\varphi(e_{11})\right]\eta_2.$$
From (\ref{alfabeta}) we conclude that $\varphi(e_{11})\eta_2=V_\eta\rho_\eta(e_{11})V_\eta^*\eta_2$ and the 
proof of sufficiency is finished.

It is easy to observe that in order to prove necessity one should repeat the same computations in the converse
direction.
\end{proof}


\begin{thebibliography}{99}
\bibitem{Alf} E.M. Alfsen and F.W. Shultz, {\it State spaces of operator algebras}, Birkhauser,
   Boston, 2001.
\bibitem{Ara} H.~Araki, Some properties of modular conjugation operator of a von Neumann
   algebra and a non-commutative Radon-Nikodym theorem with a chain rule,
   \textit{Pac. J. Math.} \textbf{50} (1974), 309--354.
\bibitem{BR} O. Bratteli and D.W. Robinson, {\it Operator Algebras and Quantum Statistical
   Mechanics I} : Second Edition,
   Springer-Verlag, New York, 1987.
\bibitem{Ch2} M.-D.~Choi, Completely positive maps on complex matrices,
   \textit{Lin. Alg. Appl.}~\textbf{10} (1975), 285--290.
\bibitem{Ch3} M.-D. Choi, Positive semidefinite biquadratic forms,
   \textit{Lin. Alg. Appl.} \textbf{12} (1975), 95--100.
\bibitem{Conn} A. Connes, Charact\'erisation des espaces vectoriels ordonn\'es
sous-jacents aux alg\`ebres de von Neumann, {\it Ann. Inst. Fourier, Grenoble}
{\bf 24} (1974), 121-155
\bibitem{Hor}
M. Horodecki, P. Horodecki, R. Horodecki, Separability of mixed states: necessary
and sufficient conditions, {\it Phys. Lett A}
{\bf 223}, 1-8 (1996)
\bibitem{Kad} R. V. Kadison, Transformations of states in operator theory and dynamics,
\textit{Topology} \textbf{3} (1965) 177-198
\bibitem{Kye} S.-H. Kye, Facial structures for the positive linear maps between matrix algebras,
\textit{Canad. Math. Bull.}~\textbf{39}(1)~(1996), 74--82.
\bibitem{LMMp} L.~E.~Labuschagne, W.~A.~Majewski and M.~Marciniak, On $k$-decomposability of
positive maps, preprint, \texttt{math-ph/0306017}.
\bibitem{LMM} L.~E.~Labuschagne, W.~A.~Majewski and M.~Marciniak, On decomposition of
positive maps, in preparation.
\bibitem{Maj} W.~A.~Majewski, Transformations between quantum states,
   \textit{Rep. Math. Phys.}~\textbf{8}~(1975), 295--307.
\bibitem {wam} W.A. Majewski, Dynamical Semigroups in the Algebraic
Formulation of Statistical Mechanics, \textit{Fortschr. Phys.} \textbf{32}(1984)1, 89--133.
\bibitem{MajO} W.~A.~Majewski, Separable and entangled states of composite
quantum systems; Rigorous description, {\it Open Systems $\&$ Information Dynamics},
{\bf 6}, 79-88 (1999).
\bibitem{MajL}
W.~A.~Majewski, ``Quantum Stochastic Dynamical Semigroup'', in \textit{Dynamics of Dissipations},
eds P. Garbaczewski and R. Olkiewicz, Lecture Notes in Physics, vol. \textbf{597}, pp. 305-316,
Springer (2002).
\bibitem{MajC}
W.~A.~Majewski, On quantum correlations and positive maps, \textit{Lett. Math. Phys.}~\textbf{67}~(2004),
125--132.
\bibitem{MM} W.~A.~Majewski and M.~Marciniak, On a characterization of positive maps,
   \textit{J.~Phys.~A: Math. Gen.}~\textbf{34}~(2001), 5863--5874.
\bibitem{MS} W.~A.~Majewski and R.~Streater, Detailed balance and quantum dynamical maps,
 \textit{J.~Phys.~A: Math. Gen.}~\textbf{31}~(1998), 7981--7995.
\bibitem{Per} A.~Peres, Separability criterion for density matrices, \textit{Phys. Rev. Lett} \textbf{77}, 1413 (1996)
\bibitem{Sto1} E.~St{\o}rmer, Positive linear maps of operator algebras,
   \textit{Acta Math.}~\textbf{110}~(1963), 233--278.
\bibitem{Sto3} E. St{\o}rmer, On the Jordan structure of $C^*$-algebras,
   \textit{Trans. Amer. Math. Soc}~\textbf{120} (1965), 438--447.
\bibitem{Sto2} E. St{\o}rmer, Decomposable positive maps on $C^*$-algebras,
   \textit{Proc. Amer. Math. Soc.} \textbf{86} (1982), 402--404.
\bibitem{Wit}
G. Wittstock, {\it Ordered Normed Tensor Products} in
``Foundations of Quantum Mechanics and Ordered Linear Spaces''
(Advanced Study Institute held in Marburg)
A. Hartk\"amper and H. Neumann eds. \textit{Lecture Notes in Physics}
vol. \textbf{29}, Springer Verlag 1974.
\bibitem{Wor} S. L. Woronowicz, Positive maps of low dimensional matrix algebras,
   \textit{Rep. Math. Phys.} \textbf{10} (1976), 165--183.
\end{thebibliography}
\end{document}